\documentclass[aps,prl,twocolumn,floatfix,showpacs]{revtex4}

\usepackage{graphics}
\usepackage{bm}

\newcommand{\myscaleboxa}[1]{\scalebox{0.45}[0.40]{#1}}
\newcommand{\myscaleboxb}[1]{\scalebox{0.45}[0.41]{#1}}
\newcommand{\myscalebox}[1]{\scalebox{0.45}[0.45]{#1}}

\newcommand{\be}{\begin{equation}}
\newcommand{\bea}{\begin{eqnarray}}
\newcommand{\eea}{\end{eqnarray}}
\newcommand{\ee}{\end{equation}}

\begin{document}

\title{Accurate retrieval of structural information from laser-induced photoelectron and high-harmonic spectra
by few-cycle laser pulses}

\author{Toru Morishita,$^{1,2}$ Anh-Thu Le,$^1$ Zhangjin Chen$^1$
and C.~D. Lin$^1$}

\affiliation{$^1$Department of Physics, Cardwell Hall, Kansas
State University, Manhattan, KS 66506, USA\\
$^2$Department of Applied Physics and Chemistry, University of
Electro-Communications, 1-5-1 Chofu-ga-oka, Chofu-shi, Tokyo,
182-8585, Japan}

\date{\today}

\begin{abstract}

By analyzing ``exact'' theoretical results from solving the
time-dependent Schr\"odinger equation of atoms in few-cycle laser
pulses, we established the general conclusion that differential
elastic scattering and photo-recombination cross sections of the
target ion with {\em free} electrons can be extracted accurately
from laser-generated high-energy electron momentum spectra and
high-order harmonic spectra, respectively. Since both electron
scattering and photoionization (the inverse of
photo-recombination) are the conventional means for interrogating
the structure of atoms and molecules, this result shows that
existing few-cycle infrared lasers can be implemented for
ultrafast imaging of transient molecules with temporal resolution
of a few femtoseconds.

\end{abstract}


\maketitle

Electron diffraction and X-ray diffraction are the conventional
methods for imaging molecules to achieve spatial resolution of
better than sub-Angstroms, but they are incapable of achieving
temporal resolutions of femto- to tens of femtoseconds, in order
to follow chemical and biological transformations. To image such
transient events, large facilities like ultrafast electron
diffraction method \cite{zewail06} and X-ray free-electron lasers
(XFELs) are being developed. Instead of pursuing these expensive
technologies,  here we provide  the needed quantitative analysis
to show that existing few-cycle infrared lasers can be implemented
for ultrafast imaging of transient molecules.

  When an atom is exposed to an infrared laser,
  the atom is first tunnel ionized with the release of an electron.
 This electron is placed in the oscillating
 electric field of the laser and may be driven back to
 revisit its parent ion. This re-encounter incurs, in the
 second step, various elastic and inelastic electron-ion
 collision phenomena where the structural information of the target is embedded.
 The possibility of using such laser-induced returning electrons
  for self-imaging molecules has been discussed frequently in the past.
  Theoretical studies of laser-induced
  electron momentum images of simple molecules do show interference
  maxima and minima typical of diffraction images, but they are
  observed only for large internuclear distances
  \cite{zou,spanner,hu,lein,yurchenko}. In addition,  for quantitative analysis, the role of
  laser fields on these diffraction images still has to be
  understood \cite{spanner}. More recently, it was reported that  the outermost molecular orbital
   of N$_2$ molecules can be extracted from the high-harmonic generation (HHG) spectra using
   the tomographic procedure \cite{itatani}. This interesting
   result has generated a lot of excitement, but the reported
   results are obtained based on  a number of assumptions \cite{hoang,patch,david07}.
  To make dynamic chemical imaging with infrared lasers as a
  practical tool, general theoretical considerations, especially the validity of the
  extraction procedure, should be examined carefully.

 In this Letter, we show that elastic scattering cross sections
 of the target ion by free electrons can be accurately extracted
 from laser-induced photoelectron momentum spectra. We also show
 that accurate photo-recombination
 cross sections of the target ion can be extracted from the HHG
 spectra. Our conclusions are based on ``exact'' theoretical
 results by solving the time-dependent Schr\"{o}dinger equations (TDSE) of
 several atoms in intense laser fields.  While our conclusions
 are derived from atomic targets, the same conclusions should
 hold for molecules too (where ``exact'' calculations are not
 possible). For molecules, the results have important implications.
 Both elastic scattering and photoionization  are the standard methods
 for studying the structure of atoms and molecules in conventional
 energy domain measurements, thus high-energy photoelectrons and
 high harmonics generated by infrared lasers offer the  promise
 for revealing the structure of the target, with the temporal
 resolution offered by the ultrashort laser pulses.

Consider a typical few-cycle laser pulse, with mean wavelength of
800 nm and peak intensity of $10^{14}$ W/cm$^2$. The electric
field $\bm{F}(t)=-\partial\bm{A}(t)/ \partial t$ and the vector
potential $\bm{A}(t)$ of such a laser pulse are depicted in Fig.
1(a). By placing a hydrogen atom in such a laser pulse, we solved
the TDSE to obtain the photoelectron energy and momentum
distributions, shown in Figs. 1(b) and 1(c), respectively. Fig.
1(d) shows the electron momentum image of Ar in the same laser
pulse. The theoretical method for solving the TDSE has been
described previously \cite{chen}.

In Fig. 1(b), two particular energies, $2 U_p$ and $10 U_p$, are
marked, where $U_p=A_0^2/4$ is the ponderomotive energy, with
$A_0$ being the peak value of the vector potential of the laser
pulse. (We use atomic units in this paper.)  These two energy
values are important, the former is the maximum energy the
electron can reach if it is released by the laser field alone,
while the latter is the maximum energy the electron will have if
it is back scattered by the parent ion \cite{paulus94}.    To
display the full electron momentum image surface in a single plot,
in Figs. 1(c) and 1(d) we normalized electron momentum
distributions   such that the total ionization yield at each
electron energy is the same. We have chosen the horizontal axis to
be along the direction of the laser's polarization and the
vertical axis along any direction perpendicular to it (due to
cylindrical symmetry of the linearly polarized light).

\begin{figure}
\mbox{\rotatebox{0}{\myscaleboxb{
\includegraphics{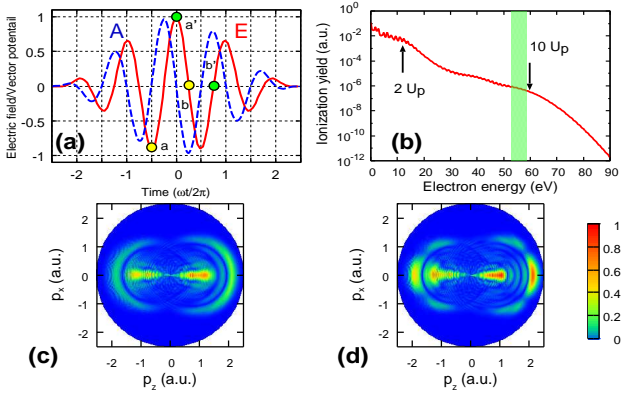}}}}
\caption{(Color online) (a) Schematic of the electric field (E)
and the vector potential (A) of a typical few-cycle pulse. (b)
Energy spectra of atomic hydrogen ionized by a 5 fs (FWHM) laser
pulse, with mean wavelength of 800 nm and peak intensity at
$10^{14}$ W/cm$^2$. (c) Normalized 2D photoelectron momentum
spectra of atomic hydrogen.
  The images are
renormalized for each photoelectron energy to reveal the global
angular distributions. (d). Momentum images of Ar in the same
pulse.} \label{fig1}
\end{figure}

  In Figs.~1(c) and 1(d), we note that in the large momentum region, each image
    exhibits two half circular rings,
  one on the ``left'' and one on the ``right'', with the center of
  each circle shifted from the origin. The rings are very
  similar for the two targets. We will call these circular rings back rescattered
  ridges (BRR), representing electrons that have been rescattered
  into the backward directions by the target ion.  The BRR on the
  ``right'' is from electrons born at time near `a' (Fig.~1(a)),
  traveling to the ``right'' and then returning to the target
  ion at time near `b',  where they are rescattered back to the
  right. Each momentum half circle is represented approximately
  by $A_r\hat{\bm{p}}_z+p_0\hat{\bm{p}}_r$, where the second term is
  the momentum of the backscattered electron and the first term is the momentum
  added to
  the electron as it propagates from `b'
    to the end of the laser pulse.
The magnitude of the momentum
  $p_0$ is related to  the ponderomotive energy by
  $3.17\overline{U}_p=p^2_0/2$ (where $\overline{U}_p=A^2_r/4$, and $A_r$ is the vector
  potential at `b' ),
  which is the maximum energy of electrons that return to revisit
  the parent ion.  For  back scattered electrons, the two momentum terms add to give high-energy
  photoelectrons, reaching a maximum of $10 U_p$ for electrons that
  have been scattered by $180^{\circ}$ \cite{paulus94}. If the electrons are
  scattered into the forward direction, the two momentum terms
  subtract from each other, resulting in lower energy electrons.
 Similar back scattered electrons are found on the ``left''.
  These are from electrons that were born near `a$^{\prime}$' and rescattered
  back to the left near `b$^{\prime}$' [Fig.1(a)]. Both the shift of the center
  and the radius are smaller due to rescattering occurring in the
  smaller vector potential near `b$^{\prime}$'.

\begin{figure}
\mbox{\rotatebox{0}{\myscaleboxa{
\includegraphics{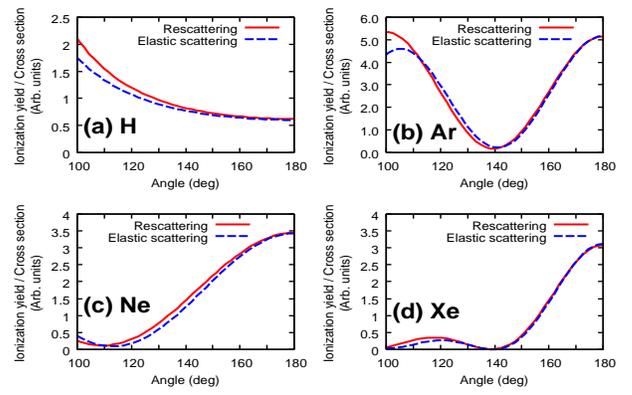}}}}
\caption{(Color online) Angular distributions of photoelectrons
along the BRR compared to the differential elastic scattering
cross
 sections of the target ion.  Each BRR  is taken to be the outer
 half circle on the right side of Fig.~1(c) and 1(d), respectively.
 (a) for H target. (b) for Ar, (c)  for Ne, (d)  for Xe.}
 \label{fig2}
\end{figure}

 In Figs. 1(c) and 1(d), we note that the yields on the BRR for H
 and Ar are quite different, where the former is monotonic and the
 latter has a clear minimum.
 Taking the actually calculated photoelectron yields (without
 the normalization as shown in the figure)   we compare the angular dependence
 of the intensities along BRR
  with the elastic differential cross
 sections of the target ion by {\em free} electrons at energy $E=p^2_0/2$.
 The results are shown in Fig. 2(a) for H target and 2(b) for Ar,
 where the scattering angles are measured from the direction
 of the ``incident'' electron beam.  Good agreement
 between the two results for each target atom can be seen. Such
 good agreement has been duplicated at different laser intensities
 and other atomic systems [see Ne and Xe in Figs.~2(c) and 2(d), respectively.].
 These results show that laser-induced
 momentum images on the BRR can be used to obtain  elastic
 scattering cross sections of free electrons by the target ion. For
atomic hydrogen, the elastic scattering cross section is given by
the Rutherford formula. For other atomic ions, minima in the
differential cross section do occur and they are due to
interference between electrons scattered by the short-range
potential and by the Coulomb potential \cite{messiah}.   We
comment that momentum spectra due to back scattered electrons have
been observed experimentally earlier \cite{yang} but there has
been no quantitative analysis.

According to the intuitive rescattering model, the photoelectron
yield along the  BRR may be interpreted as due to back scattering
of the returning electron wave packet. To test this idea, we write
the photoelectron momentum yields $I(\bm{p})$ along the BRR by
$I(\bm{p})=\sigma(p_0,\theta)F(p_0,\theta)$, with
$\bm{p}=A_r\hat{\bm{p}}_z+p_0\hat{\bm{p}}_r$, where
$\sigma(p_0,\theta)$ is the elastic differential cross section for
each ion by a free electron with energy $E=p_0^2/2$, and $\theta$
is the scattering angle of the free electron.

\begin{figure}
\mbox{\rotatebox{0}{\myscaleboxa{
\includegraphics{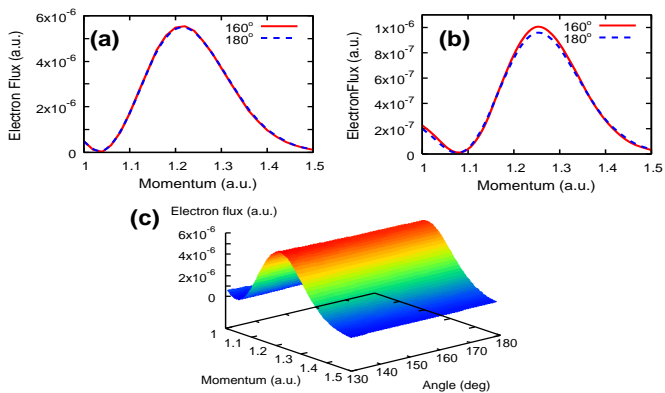}}}}
\caption{(Color online) Electron wave packets extracted from
photoelectron momentum images along the BRR are shown to be
identical for scattering angles of 160$^{\circ}$ and
180$^{\circ}$, extracted from (a) H target, (b) Ar target, using
the same ionizing laser.   (c) The extracted electron wave packet
from H target is the same over the angular range of 130$^{\circ}$
to 180$^{\circ}$.  } \label{fig3}
\end{figure}

In Fig.~3(a), we show that the extracted $F(p_0,\theta)$ from H at
two different angles. They are essentially identical such that we
may identify $F(p_0)=F(p_0,\theta)$ as the wave packet of the
returning electrons with momentum near $p_0$. Note that the wave
packet extracted from Ar target, as shown in Fig.~3(b), is
essentially identical to Fig.~3(a) except for a small shift of the
center from $p_0=1.22$ to $1.25$. The width of the wave packet is
found to be independent of the target. In Fig.~3(c) we show that
$F(p_0)=F(p_0,\theta)$ indeed holds well for electrons that have
been back scattered for angles larger than about 130$^{\circ}$.
Note that a separate electron wave packet can be retrieved from
the photoelectron momentum spectra measured on the ``left''. The
wave packet analyzed above is for a 5 fs (FWHM) pulse with the
carrier-envelope phase (CEP) $\phi=0$. The returning electron wave
packet depends on the CEP and can be used to measure the CEP of a
few-cycle pulse. For long  pulses, electrons along each side of
the BRR will exhibit oscillations characteristic of ATI peaks, and
elastic scattering cross sections by free electrons can be
extracted from the envelope of the momentum images on the BRR
\cite{chen07}.

  The above results, based on the exact solution of TDSE,
  clearly established that the electron yields on the BRR, can be
  viewed as the backscattering of the returning electron wave
  packet. According to the ``simple man'' model,
   the HHG is due to the emission of
  photons following the photo-recombination of the same returning
  electrons. Since the
   returning electron wave packets have been shown largely
   independent of target atoms for the same laser pulse, the
   difference in the HHG spectra would then be attributed to
   the recombination cross sections. To check this idea, we
    choose a companion atom to have the same ionization
   potential $I_p$ such that the cutoff energies will be identical.
    In the calculation, we compare the HHG spectra for Ne
   and scaled model hydrogen atom, generated by the same laser
   field, by solving the TDSE.
   For Ne atom, it is ionized from the 2p$_0$ state, and for H,
   we chose an effective charge such that its ground 1s state
   has the same ionization potential as Ne.  For the laser, we
   chose a 5.2 fs pulse, with mean wavelength of 1064 nm and
   intensity of $2\times 10^{14}$ W/cm$^2$. The longer wavelength was used such
   that harmonics would extend over a larger photon energy range.
   From the calculated HHG yield, we divide each by its ``exact''
   photo-recombination cross section. We compare in Fig.~4(a) the
   resulting ``electron wave packets'' (normalized) vs the HHG order.
   Clearly  the two electron wave packets are very close to each
   other, showing that the differences in the HHG spectra between
   the two targets are coming from the different photo-recombination
   cross sections. Note that  the kinetic energy of the photoelectron
   is related to the photon energy by $\hbar\omega=k^2/2+I_p$.
   Also note that the deduced
   ``electron wave packet'' shows many oscillations. To study the structural
   information of a target, it is much easier to factor out these oscillatory features
   by using a known companion atomic target.

\begin{figure}
\mbox{\rotatebox{0}{\myscalebox{
\includegraphics{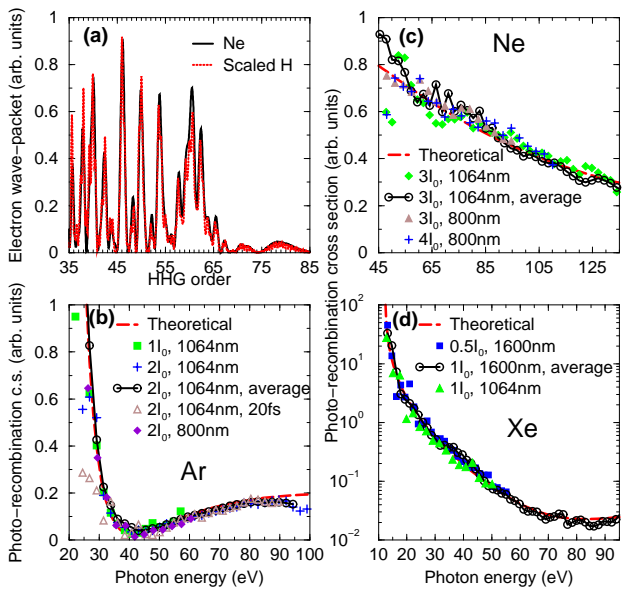}}}}
\caption{(Color online) (a) Comparison of the ``electron
wavepackets'' extracted from the HHG spectra of Ne and scaled H
generated by a 5.2 fs laser pulse with peak intensity of $2\times
10^{14}$ W/cm$^2$ and mean wavelength of 1064 nm. Extracted
photo-recombination cross sections of Ar (b), Ne (c) and Xe (d),
using different laser pulses, plotted vs the photon energy. For
each panel, the dashed red line is the ``exact'' theoretical cross
section, calculated using the same model potential, as used in the
TDSE. ($I_0=10^{14}$ W/cm$^2$.) } \label{fig4}
\end{figure}

The good agreement in the deduced ``electron wave packet'' in
Fig.~4(a) prompts us to ask whether one can obtain accurate
photo-recombination cross sections from the HHG spectra of an
unknown target by comparing it with the HHG spectra from a known
atom with identical or nearly identical ionization potential. For
our purpose we generated the HHG spectra for Ar and scaled H
atoms,  with the effective charge of H chosen to have identical
ionization potential as Ar. Different laser intensities and
wavelengths were used to generate HHG from which Ar recombination
cross sections are derived. If the procedure is valid, the
extracted recombination cross sections should be independent of
the lasers used, except for the range of photon energies covered.
In Fig.~4(b) we compare the extracted cross sections from the HHG
generated by various laser pulses with the ``exact''
photo-recombination cross section calculated for Ar. One can see
that the deduced values scatter nicely around the ``exact'' one.
The fluctuation of the extracted cross sections can be reduced if
the HHG intensity is taken  from the averaged HHG {\em amplitudes}
with nearby laser intensities. Thus the smooth black line is
obtained from averaging over eleven intensities within $\pm 5\%$
of the mean intensity of $2\times10^{14}$ W/cm$^2$. This coherent
averaging sharpens the odd harmonics and reduces the harmonic
yields in between, similar to the effect of propagation of HHG in
the medium.   Note that in the ``exact'' calculation we include
dipole transitions from 3p to both s and d continuum states. The
minimum in the cross section occurs near the Cooper minimum where
the d-wave electric dipole moment changes sign. Comparing the two
curves we can say that accurate photo-recombination cross sections
indeed can be extracted from the HHG yields. Similar test showing
good agreement has been made also on Ne and Xe atoms (Figs.~4(c)
and 4(d)), again using scaled hydrogen as the companion atoms.

Following last paragraph, we now comment the assumptions made in
Itatani {\it et al} \cite{itatani}. Their assumption that the
returning electron wave packets for Ar and N$_2$ (they have nearly
identical ionization potentials) are the same under the same laser
pulse is confirmed by our calculations which are based on exact
solution of the TDSE. However, it was also assumed
\cite{itatani,david07} that the dipole matrix elements can be
calculated using plane waves instead of ``exact'' scattering
waves. This leads to results that are quite different from ours.

 In this Letter we have identified the spectral region where the
 {\em nonlinear} laser-atom interaction can be simplified to extract
 the {\em linear} interaction between a returning electron wave packet
 with the atomic ion. Even though the results were presented for
 atomic targets, the same simplifications should hold for
 molecular targets as well. For molecules, this opens up the exciting
opportunity of using infrared lasers for ultrafast imaging of
 molecules undergoing structural transformation. Both elastic electron scattering and
 photoionization (the inverse of
 photo-recombination) processes are the well-tested means for
 probing the structure of molecules. They depend linearly on the electron current
 and photon intensity, respectively. Theoretically, these cross
 sections can be calculated very precisely. A computer package for obtaining elastic scattering cross
 section and photoionization cross section within such a model
 has been published recently by Tonzani \cite{tonzani}. By
 extracting elastic and/or photo-recombination cross sections of
 molecules using few-cycle infrared lasers, structural changes of
 the molecules can be determined with temporal resolution of a
 few femtoseconds. In conclusion, we have established the theoretical foundation for
 carrying out structural analysis of molecules with infrared
 lasers. If this roadmap is implemented experimentally,
 table-top infrared lasers would offer a very competitive
 new technology for ultrafast time-resolved chemical imaging,
 with temporal resolution down to a few femtoseconds.

This work was supported in part by the Chemical Sciences,
Geosciences and Biosciences Division, Office of Basic Energy
Sciences, Office of Science, U. S. Department of Energy. TM is
supported by a Grant-in-Aid for Scientific Research (C) from the
Ministry of Education, Culture, Sports, Science and Technology,
Japan,  by the 21st Century COE program on ``Coherent Optical
Science''.


\begin{thebibliography}{xx}

\bibitem{zewail06} A. Zewail, Annu. Rev. Phys. Chem. {\bf 57}, 65
(2006).

\bibitem{zou} T. Zuo {\it et al.}, Chem. Phys. Lett. {\bf 259}, 313 (1996).

\bibitem{spanner} M. Spanner {\it et al.},
J. Phys. B: At. Mol. Opt. Phys. {\bf 37}, L243 (2004).

\bibitem{hu}  S.~X. Hu, L.~A. Collins,  Phys.
Rev. Lett. {\bf 94}, 073004 (2005).

\bibitem{lein} M. Lein {\it et al.}, Phys.
Rev. A {\bf 66}, 051404(R) (2002).

\bibitem{yurchenko} S.~N. Yurchenko {\it et al.},  Phys. Rev. Lett. {\bf 93}, 223003
(2004).

\bibitem{itatani} J. Itatani {\it et al.}, Nature {\bf 432}, 867 (2004).

\bibitem{hoang} V.~H. Le {\it et al.}, Phys. Rev. A. {\bf 76}, 013414 (2007).

\bibitem{patch} S. Patchkovskii {\it et al.}, Phys. Rev. Lett. {\bf 97}, 123003 (2006).

\bibitem{david07} J. Levesque {\it et al.}, Phys. Rev. Lett. {\bf 98},
183903 (2007).

\bibitem{chen}  Z. Chen {\it et al.}, {\it Phys. Rev. A} {\bf 74}, 053405 (2006).

\bibitem{paulus94} G.~G. Paulus {\it et al.},
J. Phys. B: At. Mol. Opt. Phys. {\bf 27}, L703 (1994).

\bibitem{messiah} A. Messiah, {\it Quantum Mechanics.} Vol. I, p. 428,
(North-Holland Publishing, Amsterdam, 1958).

\bibitem{yang} B. Yang {\it et al.}, Phys. Rev. Lett. {\bf
71}, 3770 (1993).

\bibitem{chen07}  Z. Chen {\it et al.}, Phys. Rev. A  (submitted).

\bibitem{tonzani} S. Tonzani, Comput. Phys. Comm. {\bf 176}, 146-156 (2007).

\end{thebibliography}
\end{document}